\documentclass[journal,comsoc]{IEEEtran}
\usepackage[utf8]{inputenc}

\usepackage[hyphens]{url}
\usepackage[colorlinks=true,urlcolor=blue,linkcolor=red,citecolor=blue]{hyperref}

\usepackage{ragged2e}
\usepackage{cite}
\usepackage{amsmath,amssymb,amsfonts}
\usepackage{algorithmic}
\usepackage{graphicx}
\usepackage{textcomp}
\usepackage{soul}
\usepackage{makecell}
\usepackage{booktabs} 
\usepackage{multirow}
\usepackage{subfigure, soul}

\graphicspath{ {./Figures/} }
\usepackage[thinlines]{easytable}
\usepackage{adjustbox}
\usepackage{acronym, soul, cancel}
\usepackage{makecell}
\newcommand{\tabitem}{~~\llap{\textbullet}~~}
\usepackage[table]{xcolor}

\definecolor{bl0}{rgb}{0.36, 0.54, 0.66}
\definecolor{bl1}{rgb}{0.94, 0.97, 1.0}
\usepackage{multido}

\newcommand{\blackbullets}[1]{\multido{\i=1+1}{5}{\ifnumgreater{\i}{#1}{\color{bl1}}{\color{bl0}}\Large \textbullet\kern 0.1em}}

\definecolor{grannysmithapple}{rgb}{0.66, 0.89, 0.63}
\definecolor{green(colorwheel)(x11green)}{rgb}{0.0, 1.0, 0.0}
\definecolor{lightgreen}{rgb}{0.56, 0.93, 0.56}
\definecolor{lightcoral}{rgb}{0.94, 0.5, 0.5}

\usepackage{hyperref}

\usepackage{xspace}
\usepackage{bbding}

\makeatletter
\DeclareRobustCommand\onedot{\futurelet\@let@token\@onedot}
\def\@onedot{\ifx\@let@token.\else.\null\fi\xspace}

\def\etal{\emph{et al}\onedot}

\newacro{e2e}[E2E]{end-to-end}
\newacro{mla}[MLA]{microlens arrays}
\newacro{vvc}[VVC]{versatile video coding}
\newacro{hmd}[HMD]{head-mounted display}
\newacro{hevc}[HEVC]{high-efficiency video coding}
\newacro{odv}[ODV]{omnidirectional video}
\newacro{qoe}[QoE]{quality of experience}
\newacro{dhl}[DHL]{digital holographic microscopy}

\newacro{ssim}[SSIM]{structural similarity index measure}
\newacro{mse}[MSE]{mean squared error}

\newacro{mcts}[MCTS]{motion-constrained tile set}
\newacro{sei}[SEI]{supplemental enhancement information}
\newacro{sei}[SEI]{supplemental enhancement information}
\newacro{cmp}[CMP]{cube map projection}
\newacro{tsp}[TSP]{truncated square pyramid}
\newacro{erp}[ERP]{equirectangular projection}
\newacro{3dof}[3DoF]{three degrees of freedom}
\newacro{dof}[DoF]{degrees of freedom}
\newacro{6dof}[6DoF]{six degrees of freedom}
\newacro{uav}[UAV]{unmanned aerial vehicle}
\newacro{6dof}[6DoF]{six degrees of freedom}
\newacro{srd}[SRD]{spatial relationship description}
\newacro{omaf}[OMAF]{omnidirectional media format}
\newacro{webrtc}[WebRTC]{web real-time communication}
\newacro{isobmff}[ISOBMFF]{ISO base media file format}
\newacro{dash}[DASH]{dynamic adaptive streaming over HTTP}
\newacro{rwp}[RWP]{region-wise packing}
\newacro{avc}[AVC]{advanced video coding}
\newacro{ai}[AI]{artificial intelligence}
\newacro{co2}[CO2]{carbon dioxide}
\newacro{rwqr}[RWQR]{region-wise quality ranking}
\newacro{ivo}[IVO]{initial viewing orientation}
\newacro{rvtm}[RVTM]{recommended viewport timed metadata}
\newacro{vvc}[VVC]{versatile video coding}
\newacro{hmd}[HMD]{head-mounted display}
\newacro{hevc}[HEVC]{high-efficiency video coding}
\newacro{odv}[ODV]{omnidirectional video}
\newacro{qoe}[QoE]{quality of experience}
\newacro{mv}[MV]{multi-view}
\newacro{mcts}[MCTS]{motion-constrained tile set}
\newacro{sei}[SEI]{supplemental enhancement information}
\newacro{sei}[SEI]{supplemental enhancement information}
\newacro{cmp}[CMP]{cube map projection}
\newacro{tsp}[TSP]{truncated square pyramid}
\newacro{erp}[ERP]{equirectangular projection}
\newacro{3dof}[3DoF]{three degrees of freedom}
\newacro{6dof}[6DoF]{six degrees of freedom}
\newacro{uav}[UAV]{unmanned aerial vehicle}
\newacro{6dof}[6DoF]{six degrees of freedom}
\newacro{srd}[SRD]{spatial relationship description}
\newacro{omaf}[OMAF]{omnidirectional media format}
\newacro{webrtc}[WebRTC]{web real-time communication}
\newacro{isobmff}[ISOBMFF]{ISO base media file format}
\newacro{dash}[DASH]{dynamic adaptive streaming over HTTP}
\newacro{rwp}[RWP]{region-wise packing}
\newacro{avc}[AVC]{advanced video coding}
\newacro{dt}[DT]{digital twin}
\newacro{kpi}[KPI]{key performance indicator}
\newacro{qos}[QoS]{quality-of-service}
\newacro{qoe}[QoE]{quality-of-experience}
\newacro{qoi}[QoI]{quality-of-immersion}
\newacro{mac}[MAC]{medium access control}
\newacro{ue}[UE]{user-equipment}
\newacro{marl}[MARL]{multi-agent reinforcement learning}
\newacro{rwqr}[RWQR]{region-wise quality ranking}
\newacro{ivo}[IVO]{initial viewing orientation}
\newacro{rvtm}[RVTM]{recommended viewport timed metadata}
\newacro{3gpp}[3GPP]{3rd generation partnership project}
\newacro{dvb}[DVB]{digital video broadcast}
\newacro{mpeg}[MPEG]{motion picture experts group}
\newacro{miv}[MIV]{MPEG immersive video}
\newacro{v3c}[V3C]{visual volumetric video-based coding}
\newacro{pcc}[PCC]{point cloud coding}
\newacro{g-pcc}[G-PCC]{geometry-based \acs{pcc}}
\newacro{v-pcc}[V-PCC]{video-based \acs{pcc}}
\newacro{htc}[HTC]{holographic-type communication}
\newacro{raht}[RAHT]{region adaptive hierarchical transform}
\newacro{vps}[VPS]{video parameter set}
\newacro{mpd}[MPD]{media presentation description}
\newacro{srtp}[SRTP]{secure real-time transport protocol}
\newacro{rtp}[RTP]{real-time transport protocol}
\newacro{rtcp}[RTCP]{real-time transport control protocol}
\newacro{mec}[MEC]{multi-access edge computing}

\newacro{v-dmc}[V-DMC]{video-based dynamic mesh coding}
\newacro{afx}[AFX]{animation Framework extension}
\newacro{mlp}[MLP]{multilayer perceptron}
\newacro{ct}[CT]{cyber twin}
\newacro{pt}[PT]{physical twin}
\newacro{vr}[VR]{virtual reality}
\newacro{xr}[XR]{extended reality}
\newacro{ar}[AR]{augmented reality}
\newacro{lf}[LF]{light field}
\newacro{vrif}[VRIF]{\acs{vr} industry forum}
\newacro{nerf}[NeRF]{neural radiance field}
\newacro{nerv}[NeRV]{neural representations for videos}
\newacro{gpt3}[GPT3]{generative retrained transformer 3}
\newacro{npu}[NPU]{neural processing unit}
\newacro{fg-mg}[FG-MG]{focus group on Metaverse}
\newacro{dh}[DH]{digital hologram}
\newacro{slm}[SLM]{spatial light modulator}
\newacro{eac}[EAC]{equi-angular cubemap}
\newacro{lfd}[LFD]{light field display}
\newacro{ctc}[CTC]{common test condition}
\newacro{nrsh}[NRSH]{numerical reconstruction software for digital holograms}
\newacro{iot}[IoT]{internet of things}
\newacro{usd}[USD]{universal scene description}
\newacro{flops}[FLOPS]{floating point operations per second}
\newacro{llm}[LLM]{large language model}
\newacro{sai}[SAI]{subaperture image}
\newacro{rgb}[RGB]{red-green-blue}

\title{Immersive Media and Massive Twinning: Advancing Towards the Metaverse}

\begin{document}
\author{Wassim~Hamidouche, Lina~Bariah, and M{\'e}rouane~Debbah 


}

\maketitle

\begin{abstract}
The advent of the Metaverse concept has further expedited the evolution of haptic, tactile internet, and multimedia applications with their VR/AR/XR services, and therefore, fully-immersive sensing is most likely to define the next generation of wireless networks as a key to realize the speculative vision of the Metaverse. In this magazine, we articulate different types of media that we envision will be communicated between the cyber and physical twins in the Metaverse. In particular, we explore the advantages grasped by exploiting each kind, and we point out critical challenges pertinent to 3D data processing, coding, transporting, and rendering. We further shed light on the role of future wireless networks in delivering the anticipated quality of immersion through the reliable streaming of multimedia signals between the digital twin and its physical counterpart. Specifically, we explore emergent communication paradigms, including semantic, holographic, and goal-oriented communication, which we expect to realize energy and spectrally efficient Metaverse while ensuring ultra-low latency.
\end{abstract}

\maketitle

\begin{figure*}[h!]
\centering
  \includegraphics[width=0.92\textwidth]{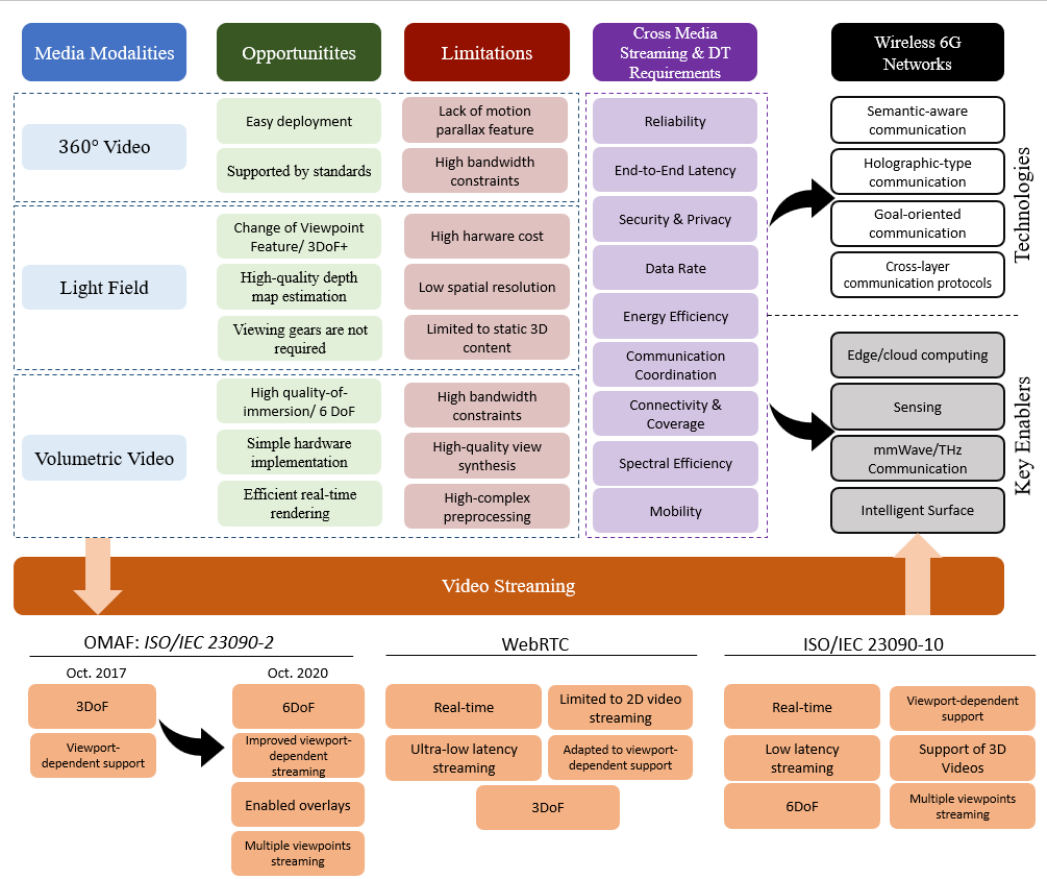}
  \caption{Immersive Media for Digital Twin: Opportunities, Limitations, 6G Technologies.}
  \label{fig:summary}
\end{figure*}

\section{Introduction} 

\color{black}
The acceleration witnessed in the maturing of the Metaverse paradigm is fueled by the emergence of the \ac{dt} technology, where the latter constitutes the cornerstone in realizing the Metaverse. The \ac{dt} can be defined as an identical digital replica of a physical environment, decoupling the static and moving objects and mimicking close-to-real interactions among them~\cite{bariah2022interplay}. On the other hand, the Metaverse is designed through enabling a massive twinning process. {In particular, the massive twinning concept implies a scenario where an extensive network of interconnected digital twins is created. Each digital twin could represent multiple real-world entities, such as geographical areas, objects, environments, etc. The interconnection of these digital twins would aim to create a comprehensive and dynamic representation of the physical world within the Metaverse. This concept aligns with the idea that the Metaverse is not just a singular virtual space but a complex and interconnected digital environment that mirrors diverse aspects of the real world.}

The Metaverse can be characterized by three main properties: immersion, interaction, and persistence. These properties are made possible through advancements in immersive audio and visual media, tactile internet, and wireless communication using 5G networks, which offer ultra-low latency and high throughput communication. Additionally, hardware improvements at various levels of the communication chain, including cloud infrastructure, edge devices, and sensors, contribute to enhanced computing capabilities. The immersion and interaction properties of the Metaverse enable users to engage with virtual environments and experience a seamless blend of real and virtual information. This continuous interaction and development within the \ac{xr} environment promote innovation and new applications across consumer, enterprise, and industrial sectors. 

The potential economic impact of the Metaverse is significant, with the total addressable market projected to reach between \$8 trillion and \$13 trillion by 2030\footnote{{Metaverse and Money Decrypting the Future. Citi Global Perspectives \& Solutions,} \href{https://www.citigroup.com/global/insights/citigps/metaverse-and-money_20220330}{Link} Accessed: 12-13-2023.} A recent market study conducted by Deloitte in 2023 suggests that Europe alone could benefit from a Metaverse economy of \$259 billion to \$489 billion by 2035\footnote{{The Metaverse does not  need virtual reality or web3,  but they may help. Deloitte,}  \href{https://www2.deloitte.com/content/dam/Deloitte/uk/Documents/technology-media-telecommunications/deloitte-uk-digital-consumer-trends-2022-metaverse.pdf}{Link} Accessed: 12-13-2023.}. Furthermore, Gartner, a leading technological research and consulting firm, predicts that 25\% of people will spend at least one hour per day on the Metaverse platform for personal and professional purposes\footnote{{Gartner, Press Release}. \href{https://www.gartner.com/en/newsroom/press-releases/2022-02-07-gartner-predicts-25-percent-of-people-will-spend-at-least-one-hour-per-day-in-the-metaverse-by-2026}{Link}}.

\color{black}

\begin{table*} 
\caption{State-of-the-art papers discussing Metaverse concepts and challenges} \label{tab:soa}
\centering
\begin{tabular}{r|c|l} 
   \toprule
  Reference & Year & Topics covered \\
      \midrule
Lee \etal \cite{lee2021all} & 2021 &
     \thead[l]{  \tabitem Enabling technologies for the Metaverse, e.g., AI, blockchain, computer vision, IoT, edge-computing.  \\ 
       \tabitem Creating an ecosystem at the Metaverse: avatars, content creation, virtual economy,  \\ social acceptability, etc.  }
      \\ 
       \midrule
 Wang \etal \cite{wang2022survey}  & 2022 &
   \thead[l]{    \tabitem Architecture, enabling technologies, and main features of the Metaverse.  \\
       \tabitem Security and privacy threats in the Metaverse.}
    \\ 
  \midrule
  Xu \etal \cite{xu2022full} & 2022 &
    \thead[l]{   \tabitem Metaverse architecture.  \\
       \tabitem Networking and communications challenges. \\
       \tabitem Cloud and edge computational challenges. \\
       \tabitem The interplay of blockchain and the Metaverse.}
  \\ 
  \midrule
  Dwivedi \etal \cite{dwivedi2022metaverse} & 2022 &
     \thead[l]{  \tabitem Overview, applications, and limitations of the Metaverse.  \\
       \tabitem Different aspects of the Metaverse, including legal, human, marketing, advertisement, retail, etc. }
   \\ 
  \midrule
 Yang \etal \cite{yang2022fusing} &  2022 &
     \thead[l]{  \tabitem AI in the Metaverse.  \\
       \tabitem Blockchain in the Metaverse. }
 \\ 
  \midrule
Khan \etal \cite{khan2022metaverse} & 2022 &
      \thead[l]{ \tabitem General architecture for Metaverse-enabled wireless networks. \\
       \tabitem Driving applications, trends, and enabling technologies.  }
   \\ 
  \midrule
Tang  \etal \cite{tang2022roadmap} & 2022 &
    \thead[l]{ \tabitem Wireless communication and networking at the Metaverse: integrated terrestrial and  \\ non-terrestrial networks, mmWave, edge-computing, integrated sensing and communication, etc.   }
  \\ 
  \midrule
Wang  \etal \cite{wang2022mobile} & 2022 &
      \tabitem 6G computing paradigms for the Metaverse.
      \\   
 \midrule
Ours   & 2023 & \thead[l]{  \tabitem 3D visual content modalities for the Metaverse.  \\
\tabitem Standards for compression and transport of 3D visual content. \\
\tabitem Enabling wireless technologies and paradigms for the Metaverse. \\
\tabitem Open challenges on both 3D visual content streaming, and low latency and high throughput  \\ wireless communication. } \\
  \bottomrule
\end{tabular}
\end{table*}

\color{black}

The Metaverse relies on the seamless integration of real and virtual realms, where various forms of multimedia data are exchanged. The increasing availability of virtual and augmented reality devices, along with advancements in computing resources and high-resolution equipment, has paved the way for immersive communication as a means of accessing the Metaverse. Immersive communication involves the exchange of multimedia data and natural haptic signals with remote devices. The effectiveness of immersive communication depends on the ability of wireless nodes to interact with remote environments and accurately perceive and quantify this interaction using all senses. To achieve the desired \ac{qoe}, it is crucial that participating nodes in remote environments have high-resolution haptic and multimedia perception. In order to attain ultimate immersion, future wireless networks must adhere to several \acp{kpi}, including low communication latency (sub-millisecond) and high throughput. These \acp{kpi} ensure a high-quality experience for multimedia data, considering factors such as spatial resolution, color depth, dynamic range, frame-rate, and glass-to-glass latency. Consequently, it is necessary to reevaluate existing wireless techniques and develop novel communication protocols capable of handling various types of multimedia data to deliver the required level of immersion for a high-fidelity Metaverse.

To this extend, in this article, we identified the different types of multimedia that are envisioned to be communicated in the era of the Metaverse, and then discussed the potential wireless schemes that manifest themselves as enablers for immersive media transfer.
This paper aims to provide an overview of different types of multimedia data that contribute to the realization of a fully-immersive Metaverse. The requirements of each type of multimedia are elucidated, highlighting how they can enhance the overall immersion. Furthermore, the latest advancements in compression and transport protocols for efficient transmission and storage of 3D visual modalities are presented. The article also delves into the wireless technologies that play a pivotal role in enabling reliable, fast, and energy-efficient multimedia communication. Fig.~\ref{fig:summary} gives a comprehensive summary of the main topics covered in this article. While numerous papers have discussed the Metaverse from various perspectives, to the best of our knowledge, this article is the first to explore the immersive media within the context of wireless networks. 
Table \ref{tab:soa} serves as a comprehensive summary of the existing related literature in the field. In this paper, we make the following key contributions:
\begin{itemize}
\item Present visual 3D media modalities enabling immersive quality of experience, including 360$^\circ$, point cloud, light field, and volumetric video.   
\item Review encoding standards and transport protocols enabling efficient transmission and storage of 3D visual content. 
\item Investigate enabling wireless technologies for reliable, real-time, and energy-efficient multimedia communication. 
\item Assess the coding efficiency and decoding latency of conventional and AI-based video encoding of volumetric video and dynamic point clouds.   
\item  Discuss open challenges and future research directions for the development and the massive deployment of the Metaverse. 
\end{itemize} \color{black}
The rest of this paper is organized as follows. Section~\ref{sec:opencha} overview the different types of media technologies, with emphasis on their basic principles, coding schemes, and standardization efforts. Section~\ref{sec:meth} puts a forward-looking vision on the enabling wireless technologies and paradigms anticipated to be employed for successfully transmitting immersive media data. Further, Section~\ref{sec:opencha} identifies several limitations and challenges encountered in immersive media communications towards the \ac{dt} and the Metaverse. Finally, Section~\ref{sec:con} concludes the paper. 

\begin{table*}[t!]
    \centering
    \caption{Acquisition, display, enabled quality of experience and used coding standards for different visual media modalities.}
    \label{tab:3Dmod}
    \begin{adjustbox}{max width=\textwidth}
    \begin{tabular}{l|c|c|c|c}
    \toprule
     {\bf Modality } &  { \bf Acquisition} & { \bf Compression standards \& formats}  & { \bf Display}   &   { \bf Visual experience } \\ \midrule
    \multirow{ 2}{*}{{\bf 360$^\circ$ video}}  & \multirow{ 2}{*}{Multiple 2D or fish-eye cameras} &  \acs{avc}/H.264, \acs{hevc}/H.265,  &  
  \multirow{ 2}{*}{2D displays or \acs{hmd}} & \cellcolor{lightcoral} \\  
             & & \acs{vvc}/H.266, VP9 and AV1 &   &  \cellcolor{lightcoral} 
 \multirow{-2}{*}{\acs{3dof}} \\  \midrule
        {\bf \Acl{lf}} & Plenoptic camera or array of cameras   & JPEG Pleno, \acs{mv}/3D-\acs{hevc}, \acs{miv}  & 2D displays, \acs{hmd}, or \acs{lfd} & \cellcolor{yellow}  \acs{3dof}+  \\  \midrule
      {\bf  Point clouds } & Multiple 2D cameras and depth sensors & JPEG Pleno, V-\acs{pcc}, G-\acs{pcc}, Draco,  Corto &  2D displays or \acs{hmd} &  \cellcolor{lightgreen} \acs{6dof} \\   \midrule
       {\bf Mesh} & Multiple 2D cameras and depth sensors & Draco, Corto, \acs{v-dmc}  & 2D displays or \acs{hmd}  &  \cellcolor{lightgreen} \acs{6dof}   \\  \midrule
       \multirow{2}{*}{{\bf \Acl{dh}}}  & \acs{dhl}, holographic interferometry  &\multirow{2}{*}{ JPEG Pleno } &  \multirow{2}{*}{ 2D displays, \acs{hmd} or \acs{slm}}  &  \cellcolor{lightgreen}   \\ 
        & computer-generated hologram synthesis methods & &  & \cellcolor{lightgreen} \multirow{-2}{*}{ \acs{6dof}}\\
        \bottomrule
    \end{tabular}
   \end{adjustbox}
         {\begin{flushleft}
 LFD: Light field display. DHL: Digital holographic microscopy. PCC: point cloud coding.   
   \end{flushleft}}
\end{table*}

\section{Immersive Media Streaming}
\label{sec:rel}
The Metaverse may benefit from a realistic representation of 3D natural scenes in high quality, especially for close-to-real visualization of objects and humans. In this section, we overview visual media modalities developed to enable high-quality streaming of 3D scenes from acquisition to display, emphasizing on standardization efforts to ensure the interoperability of devices for immersive services. Table~\ref{tab:3Dmod} summarizes the acquisition technologies, compression standards and formats, display, and enabled \ac{dof} by these visual 3D modalities.  
\subsection{3D visual signal}
Various modalities, such as 360$^\circ$ imagery, \ac{lf} data, volumetric visual signals, and \ac{dh}, have the capability to represent both natural and synthetic 3D scenes. This section provides a detailed description of these modalities, emphasizing their advantages and disadvantages when employed to depict static or dynamic 3D scenes.   \\
{\bf Omnidirectional visual signal}. An omnidirectional visual signal is presented in a spherical space with angular coordinates: the azimuth angle $\phi \in [ \pi, - \pi ]$, and the elevation or polar angle $\theta \in [ 0, \pi ]$, assuming a unit sphere (radius $r=1$) for acquisition and rendering. The sphere's origin represents the viewing reference that captures the light coming from all directions. The omnidirectional image allows the user to visualize the scene in \ac{3dof} by interacting with the scene through \ac{hmd} with head rotations: roll, yaw, and pitch. In practice, an omnidirectional visual signal is acquired by a multi-view wide-angle capture. The wide-angle capture relies, for instance, on fish-eye lenses. One fish-eye camera enables only a partial sphere capture, while multiple fish-eye camera acquisitions are combined to cover the whole sphere. This operation is performed by {\it stitching} images from different cameras into the sphere. The omnidirectional visual signal in spherical representation is then mapped into a 2D texture signal in the pre-processing stage before being encoded by conventional 2D video coding standards. \Ac{erp} is the most commonly used bijective mapping technique, particularly adapted for production. Other mapping techniques, such as \ac{cmp} and \ac{tsp}, achieve a more efficient coding estimated respectively to 25\% and 80\%~\cite{360meta} superior to \ac{erp}, and thus are more suitable for distribution\footnote{360$^\circ$ projection methods: \url{https://map-projections.net/}}.

\begin{figure}[h!]
\includegraphics[width=0.4\textwidth]{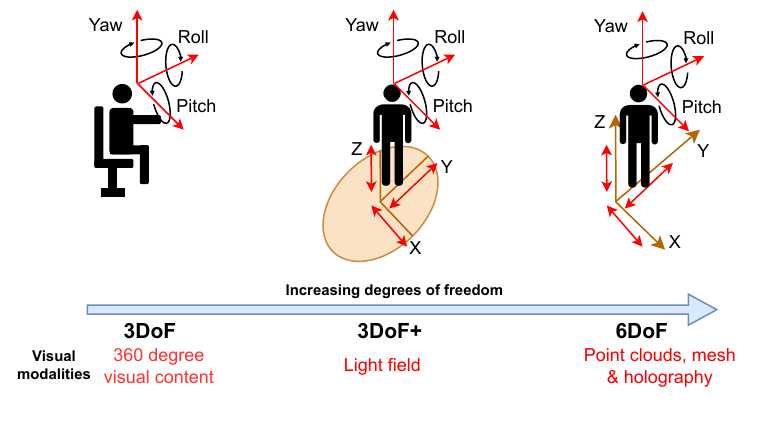}
\caption{Viewing degrees of freedom (DoF) enabled by 3D visual modalities.}
\label{fig:DoFs}
\end{figure}

One of the main limitations of \ac{odv} is the absence of the motion parallax feature. Motion parallax refers to the relative position of objects changing based on the viewer's perspective in relation to those objects. This limitation can result in discomfort and motion sickness for users. To address this limitation, \ac{lf} and volumetric visual presentations have emerged as alternative modalities. These modalities offer a visual experience comparable to the real world, incorporating up to \ac{6dof} capabilities that enable users to freely navigate through the scene, as illustrated in Fig.~\ref{fig:DoFs}. 
\\\\ 
{\bf Light field}. 
The \acl{lf} camera, also referred to as a plenoptic camera, captures both the intensity and direction of rays within a scene. These rays are described by a seven-dimensional (7D) plenoptic function, including the 3D coordinates $(x, y, z)$ of the camera position, two angles ($\theta, \phi$) for the viewing direction, wavelength $\lambda$, and time $t$. To reduce this 7D function, the time dimension is typically sampled according to the capture device's frame-rate, and the wavelength dimension is mapped to the three \ac{rgb} color components, resulting in a 5D function. In practice, each ray's intersection is determined by its position on two parallel planes, denoted as $(a, b)$ and $(u, v)$ for spatial and angular coordinates, respectively. The resulting 4D \ac{lf} function $L(a, b, u, v)$ comprises a collection of perspective images of the $(a, b)$ plane captured from different positions on the $(u, v)$ plane. There are two main approaches to \ac{lf} acquisition: camera arrays and plenoptic cameras. Plenoptic cameras, or narrow-baseline \ac{lf} acquisition systems, involve adding \ac{mla} to conventional 2D cameras. The spatial resolution of the resulting micro-image depends on the number of microlenses used, while the angular resolution is determined by the number of pixels behind each microlens. Thus, the camera sensor's full resolution is shared between spatial and angular resolutions. The micro-image can be de-multiplexed to form \acp{sai}, which group pixels with the same relative position in the microlens image. On the other hand, camera array-based \ac{lf} acquisition systems, or wide-baseline \ac{lf}, utilize camera arrays arranged in various geometries (such as a plane or a sphere at regular intervals). Each camera represents an angular sample, and the images provide spatial samples. The specific geometry employed distinguishes \ac{lf} from general multi-view capture~\cite{broxton2020immersive}. With known spatial and temporal positions between different cameras in \ac{lf} acquisition, the underlying positions and directions of rays in space and time can be derived.
{The captured light field image enables various features, including viewpoint changes and high-quality depth map estimation. 
\\\\
{\bf Volumetric video}. 
Volumetric video encompasses two media modalities: point clouds and polygonal mesh. The acquisition of volumetric video typically involves a considerable number of cameras (ranging from 10 to more than sixty) placed around the scene\footnote{Volumetric capture companies: \url{http://volumetric-video.com/volumetric-capture-companies/}}. In addition to \ac{rgb} cameras, active depth sensors are utilized to capture the geometry of the scene. Various modules process the acquired data to construct the final 3D representation of the scene, which can then be rendered at the receiver. The initial stage of pre-processing involves transforming the input data from the different camera streams into point clouds. Point clouds represent the 3D scene using unstructured points in 3D space along with their associated attributes, such as texture, reflectance, transparency, and surface normals. Color adaptation is performed to ensure consistent color distribution across the different camera streams. Next, objects within the scene are separated from the background to focus the processing on the relevant objects. Depth estimation is then carried out using a stereo camera pair. The resulting depth maps, including both predicted depth maps and depth maps obtained from \ac{rgb}-D sensors, are transformed into 3D space and combined to generate a cohesive 3D point cloud. Additional processing steps can be applied to the point cloud, such as outlier removal through cleaning techniques. The point cloud representation of the 3D scene is subsequently encoded and transmitted to end users. Alternatively, in a different scenario, the point cloud representation is first converted into a consistent mesh representation before being encoded and streamed. The use of a mesh representation offers greater compatibility with existing players and hardware decoder implementations, enabling real-time rendering on various devices.
\begin{table*}[t!]
    \centering
    \caption{Key performance of emerging codecs for 3D visual contents.}
    \label{tab:3Dcodecperf}
    \begin{adjustbox}{max width=\textwidth}
    \begin{tabular}{l|c|c|c|c|c|c|c}
    \toprule
     {\bf Codec } &  { \bf Standard }   &   { \bf \acs{mv} } &   { \bf \acs{mv} +  depth}  &   { \bf Point cloud } &   { \bf Mesh } & { \bf Compression efficiency}  & { \bf Decoding speed}  \\ \midrule
      \bf  \acs{mv}-\acs{hevc}   & \cellcolor{lightgreen} \acs{hevc}/H.265 v2  (2014) &  \cellcolor{lightgreen} \CheckmarkBold &  \cellcolor{lightcoral} \XSolidBrush & \cellcolor{lightcoral} \XSolidBrush & \cellcolor{lightcoral} \XSolidBrush & \blackbullets{2} &  \blackbullets{2} \\  \midrule
     \bf 3D-\acs{hevc}   &\cellcolor{lightgreen} \acs{hevc}/H.265 v3 (2015) &   \cellcolor{lightgreen} \CheckmarkBold &  \cellcolor{lightgreen} \CheckmarkBold  & \cellcolor{lightcoral} \XSolidBrush & \cellcolor{lightcoral} \XSolidBrush &  \blackbullets{2} &  \blackbullets{2}  \\  \midrule
    \bf \acs{miv}  & \cellcolor{lightgreen} ISO/IEC 23090-12 (2021) &  \cellcolor{lightgreen} \CheckmarkBold &  \cellcolor{lightgreen} \CheckmarkBold  & \cellcolor{lightcoral} \XSolidBrush & \cellcolor{lightcoral} \XSolidBrush & \blackbullets{4} &  \blackbullets{4} \\  \midrule
    \bf \acs{v-pcc}  & \cellcolor{lightgreen} ISO/IEC 23090-5 (2020)  &  \cellcolor{lightcoral} \XSolidBrush  &\cellcolor{lightcoral} \XSolidBrush  &  \cellcolor{lightgreen} \CheckmarkBold & \cellcolor{lightcoral} \XSolidBrush  &\blackbullets{4} &  \blackbullets{4} \\  \midrule
    \bf \acs{g-pcc}   &\cellcolor{lightgreen} ISO/IEC 23090-9 (2021) &  \cellcolor{lightcoral} \XSolidBrush &  \cellcolor{lightcoral} \XSolidBrush & \cellcolor{lightgreen} \CheckmarkBold &  \cellcolor{lightcoral} \XSolidBrush &\blackbullets{4} &  \blackbullets{3} \\ \midrule
    \bf \acs{v-dmc}  & \cellcolor{lightgreen} ISO/IEC 23090-29  &  \cellcolor{lightcoral} \XSolidBrush  & \cellcolor{lightcoral} \XSolidBrush  &   \cellcolor{lightcoral} \XSolidBrush  & \cellcolor{lightgreen} \CheckmarkBold & \blackbullets{4} &  \blackbullets{4} \\  \midrule
    \bf Draco  &   \cellcolor{lightcoral} \XSolidBrush & \cellcolor{lightcoral} \XSolidBrush &  \cellcolor{lightcoral} \XSolidBrush & \cellcolor{lightgreen} \CheckmarkBold   & \cellcolor{lightgreen} \CheckmarkBold &\blackbullets{1} &  \blackbullets{5} \\  \midrule  
    \bf Corto  & \cellcolor{lightcoral} \XSolidBrush   & \cellcolor{lightcoral} \XSolidBrush &  \cellcolor{lightcoral} \XSolidBrush & \cellcolor{lightgreen} \CheckmarkBold   & \cellcolor{lightgreen} \CheckmarkBold & \blackbullets{1} &  \blackbullets{5} \\   
        \midrule
         \bf NeRF$^\star$ (\acs{ai}) & \cellcolor{lightcoral} \XSolidBrush & \cellcolor{lightgreen} \CheckmarkBold  & \cellcolor{lightcoral} \XSolidBrush &  \cellcolor{lightcoral} \XSolidBrush & \cellcolor{lightcoral} \XSolidBrush  & \blackbullets{4} & \blackbullets{4} \\  \bottomrule
    \end{tabular}
   \end{adjustbox}
         {\begin{flushleft}
 MV: Multi-View.  ~ ~ ~ Performance metrics: High $\equiv$ \blackbullets{5}, Low $\equiv$ \blackbullets{1}.  $^\star$ {Neural radiance field (NeRF) is a fully-connected neural network that can generate novel views of complex 3D scenes, based on a partial set of 2D images, enabling continuous representation of the 3D scene.} 
   \end{flushleft}}
   \vspace{-5mm}
\end{table*}
\subsection{Visual 3D visual signal coding}
One common limitation of all 3D visual media modalities is the substantial amount of data required for their digital representation. As a result, compression techniques are crucial for efficient storage and transmission, particularly over limited bandwidth wireless channels. This section provides a comprehensive review of the current coding standards available for encoding 3D visual content. \\\\    
{\bf Conventional 2D video standards}.
In practice, 2D image and video coding standards are commonly employed to encode 3D content. The first step involves projecting the 3D content onto one or several 2D planes during the pre-processing stage. These 2D planes are then encoded using conventional video standards. By utilizing well-established 2D video standards such as \acs{avc}/H.264, \acs{hevc}/H.265, \acs{vvc}/H.266, and, AV1, and VP9 video formats, efficient and real-time coding can be achieved with existing hardware encoders. Moreover, compliant hardware decoders are widely supported by embedded devices, televisions, and web browsers. Notably, for \ac{odv} content, a 2D standard is employed after projecting the sphere onto a 2D plane. Furthermore, tailored coding tools specific to \ac{odv} have been integrated into \acs{hevc}/H.265 and \acs{vvc}/H.266, enhancing coding efficiency and enabling advanced streaming features.

Several alternative codecs, including Draco and Corto, have been specifically designed to encode 3D content in point cloud and mesh representations. These codecs offer efficient coding tools and low-complexity decoder implementations, which are widely supported by web browsers. However, the temporal redundancy present in dynamic 3D content remains untapped, as frames are typically encoded independently. Instead, \acs{hevc}/H.265 extensions including \ac{mv}-\acs{hevc} and 3D-\acs{hevc} \cite{7258339} were proposed to encode 3D video content in \ac{mv} and \ac{mv} plus {depth} representations, respectively. 
 Despite leveraging temporal redundancy, these extensions have not gained widespread industry adoption primarily due to their limited coding efficiency and high decoding complexity, which scales linearly with the number of views. 
 To overcome these shortcomings, the \ac{v3c}~\cite{10.3389/frsip.2022.883943} encompasses a group of standards (ISO/IEC 23090-xx) to encode, store and transport volumetric visual content efficiently.  \\\\
{\bf  \Acl{v3c}}. Several compression standards have been developed under the \ac{v3c}~\cite{10.3389/frsip.2022.883943}, including \ac{miv} (ISO/IEC 23090-12)~\cite{9374648}, \ac{v-pcc} (ISO/IEC 23090-5) and \ac{g-pcc} (ISO/IEC 23090-9)~\cite{8571288}. These three standards are briefly described in the following. The \ac{v-dmc} (ISO/IEC 23090-29) standard is under development and will be discussed in Section~\ref{sec:opencha}.      \vspace{-1mm} 
\begin{enumerate}
\item  The ISO/IEC \ac{miv} standard, released in 2021, was developed to support efficient coding of 3D representation of natural or synthetic-generated 3D scene captured by multiple cameras, enabling up to \ac{6dof} viewing experience. The \ac{miv} standard specifies four profiles according to the input representation to encode. The {\it baseline profile} encodes input formats including both texture and depth, while the {\it extended profile} considers, in addition to depth and texture, transparency and occupancy information. Furthermore, the {\it extended restricted sub-profile} takes a multi-plane image (MIP) with texture and transparency as input. Finally, the {\it geometry absent} profile considers inputs only with texture information. In this latter profile, the receiver can predict the depth information from the decoded texture views. 

{The \ac{miv} encoder creates a collection of depth and attribute atlases, essentially composed of patches extracted from the input views. This approach aims to generate compact representations of depth and texture input views, effectively minimizing pixel redundancy.} In addition, the encoder generates metadata that describes these atlases. Then, the \ac{miv} encoder encodes the depth and attribute atlases with a conventional video standard and the metadata according to the \ac{miv} standard specification. A decoder compliant with the \ac{miv} standard will be able to parse the bitstream, decode the texture and depth atlases with a conventional video decoder and then reconstruct the 3D visual signal based on the decoded metadata.      

\item The ISO/IEC \ac{v-pcc} standard, finalized in 2020, enables efficient coding of a dense static or dynamic point cloud. The \ac{v-pcc} standard follows a projection-based approach that first projects the point cloud representation into texture, depth, and occupancy 2D maps, which are then compressed with a conventional video standard. The projection operation is not part of the standard, and the encoder implementation may use a custom projection solution that will impact the coding efficiency. A decoder compliant with the \ac{v-pcc} standard parses the bitstream and then decodes the three 2D maps, used then to recover a reconstructed version of the input point cloud\footnote{Real-time V-PCC decoder on mobiles phones: \url{https://github.com/nokiatech/vpcc}}.

\item The ISO/IEC \ac{g-pcc} standard, released in 2021, specifies tools to encode the geometry, which is particularly convenient for sparse point clouds. The \ac{g-pcc} encoder performs sequential encoding of the geometry, then the related attributes, enabling to leverage off the decoded geometry for encoding the attributes. The real value of 3D coordinates (i.e., geometry) is first quantized into integer representations on $M$ bits per coordinate { and then represented in cube format (i.e., voxel). The voxel geometry is analyzed by whether octree (i.e., tree data structure in which each internal node has exactly eight children) or triangle
soup (i.e., triangle representation with three vertices) scheme in a second stage}. In the case of sparse point clouds, approximately only 1\% of the voxels are occupied, which makes the octree very convenient for a compact representation of the geometry. Finally, an arithmetic encoder performs lossless compression of the resulting geometry structure, exploiting the statistical correlation between neighbor points. Regarding the attributes, after an optional conversion from \ac{rgb} to YCbCr color space, the attribute is processed by one of the three available transform tools, i.e.,  \ac{raht}, predicting transform, or lifting transform. Finally, the resulting residuals are quantized and arithmetically encoded to form the \ac{g-pcc} bitstream. 
\end{enumerate}

\noindent { \color{black} {\bf Learning-based coding.} The end-to-end trainable model for image compression (a.k.a, learned codecs) can be classified into implicit and explicit coding. The explicit coding relies on variational auto-encoder transforms that project the input data into a more compact latent representation, which is then encoded by an arithmetic encoder following a distribution model. In addition, hyper-priors (i.e., mean and standard deviation) of the latent representation are also encoded with a hyperprior auto-encoder to capture their spatial dependencies by the arithmetic encoder effectively. Learned codecs initially proposed for 2D images are then extended to 2D videos and 3D visual modalities, including light field and point cloud, showing outstanding coding efficiency superior to conventional codecs.
More recently, implicit neural coding has emerged as a promising solution where a \ac{mlp} learns the mapping from 2D coordinates to \ac{rgb} colors. First, the input coordinates are mapped into high dimensional space, typically using a sequence of sine and cosine functions (frequency encoding) of dimension $L \in \mathbb{R}$. Then, the \acp{mlp} learn the 
3D density along the 5D light field of a given 3D scene (\acs{nerf}). Subsequently, the network is trained to minimize the distortion, which is usually a tradeoff between \ac{mse} and \ac{ssim}. Finally, the compression is performed by processing the \acp{mlp}' weights with three operations: pruning, quantization, and entropy coding as illustrated in Fig.~\ref{fig:implicit} to generate the bitstream.} 
\begin{figure}[h]
\includegraphics[width=0.5\textwidth]{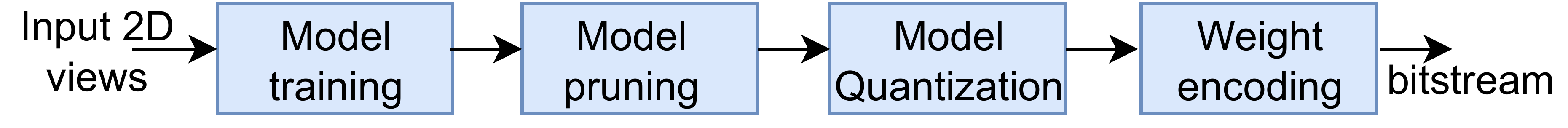}
 \caption{Implicit-based coding of 2D views for continuous 3D representation of the scene.}
  \label{fig:implicit}
\end{figure}

{Table~\ref{tab:3Dcodecperf} gives the key performance of emerging codecs for 3D visual contents. For point cloud and mesh
modalities, several studies have reported superior coding efficiency of V-\ac{pcc} compared to G-\ac{pcc}, which is in turn
superior to Draco and Corto, which do not leverage temporal redundancies. Furthermore, \ac{miv} demonstrated comparable coding performance compared to V-\ac{pcc}. For \ac{mv} modalities, \ac{nerf} showed superior performance compared to \ac{hevc}-based codecs, which are tailored for motion correlations and thus fail to capture disparity correlations. In terms of complexity, \ac{hevc}-based codecs (MV-\ac{hevc} and 3D-\ac{hevc}) can capitalize on the existing
hardware encoders and decoders of the \ac{hevc} standard. Nonetheless, the complexity experiences linear growth with the number of views. In the case of \ac{miv} and V-\ac{pcc}, which also depend on \ac{hevc}, the intricacy is primarily associated with the pre-processing and post-processing stages necessary for constructing the atlases and subsequently recovering the decoded views. Specifically, \ac{hevc} encoding is
applied solely to a select few atlas images that encompass several views. Draco and Corto decoders can run in real-time on embedded devices since they rely on simple Intra algorithms (Temporal redundancies are not considered). Finally, \ac{nerf}-based solutions can also benefit from \acp{npu} and fast conversion to mesh representation to run in real-time on mobile phones.}
\subsection{3D video streaming}
For streaming 3D content in live and offline use cases, different transport protocols can be used according to the application requirements regarding glass-to-glass latency, visual quality, and supported features. This section will focus on two widely used streaming protocols, namely \ac{omaf} \cite{9380215} and \ac{webrtc}~\cite{7160422} along with the ISO/IEC 23090-10 standard~\cite{10.3389/frsip.2022.883943} that specifically describes how to store and deliver \ac{v3c} compressed volumetric video content.  Fig.~\ref{fig:standards} illustrates the timeline of ITU-T and IEC/ISO standards specifically designed for coding and carriage of 3D visual content. \\\\     
{\bf \ac{omaf}}. The ISO/IEC 23090-2 \ac{omaf} is a \acs{mpeg} system standard developed to ensure the interoperability of devices and services targeting storage and streaming of omnidirectional media, including 360$^\circ$ images and video, spatial audio, and associated text. The first version of the standard, finalized in October 2017, provides the basic tools for streaming 360$^\circ$ images and video, enabling \ac{3dof} viewing experience. Additional tools have been integrated into the second version of the standard, released in October 2020, for more advanced features such as enhancing the viewport-dependent streaming, enabling overlays, and multiple viewpoints streaming as the first step towards \ac{6dof} viewing experience. The \ac{omaf} specifications fall within three main modules: content authoring, delivery, and player. These specifications are extensions to the \ac{isobmff} and \ac{dash}, ensuring backward compatibility with conventional 2D media formats.  
 \\\\
{\bf WebRTC}. WebRTC is an open-source framework designed for real-time and low-latency video transmission. At the WebRTC transmitter, the {\it video collector} module encodes the video and encapsulates the encoded video frames in {\ac{rtp}} packets, which are then transmitted through the \ac{srtp}. The receiver collects information on the received {\ac{rtp}} packets and sends back information to the {\it video collector} in the transport-wide feedback message of the \ac{rtcp}. Based on these control messages, the {\it bandwidth controller} module of the {\it video collector} computes network metrics such as inter-packet delay variation, queuing delay, and packet loss. These metrics are then exploited to calculate the target bit rate used by the rate control module of the video encoder that adapts the encoding parameters (quantization parameter, resolution, etc.) according to the target bit rate. However, vanilla WebRTC does not specify tools for transmitting immersive video, limiting its usage to 2D video. Nevertheless,  WebRTC was widely adopted for real-time and low latency \ac{odv} transmission by considering the 360$^\circ$ video representation as a traditional 2D video. In addition, viewport-dependent can also be supported by mixing high-resolution tiles and low-resolution 360$^\circ$ video for efficient bandwidth usage while ensuring high quality in the field of view area along with ultra-low motion-to-photon latency\footnote{Intel Advanced 360$^\circ$ Video: \url{https://www.intel.com/content/dam/www/central-libraries/us/en/documents/advanced-360video-implementation-summary-final.pdf}}.
\begin{figure}[t]
  \includegraphics[width=0.49\textwidth]{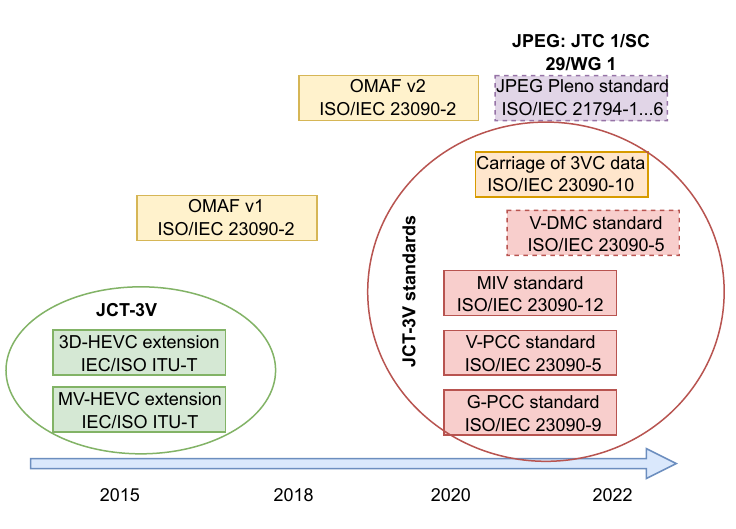}
 \caption{ITU-T and ISO/IEC standards for compression and carriage of 3D visual modalities. Dashed correspond to ongoing standardization activities.} \vspace{-3mm}
  \label{fig:standards}
\end{figure}       

\begin{table*}
\caption{Inter- and intra-twin communication challenges.} 
\label{Tab:inter-intra} 
\begin{adjustbox}{max width=\textwidth} 
\begin{tabular}{l|p{25em}|p{25em}} 
   \toprule
Requirement & \textbf{Inter-Twin} & \textbf{Intra-Twin}\\ \midrule
Communication coordination & \begin{itemize}
    \item New challenges, pertinent to coordinating the communications among an enormous number of distributed \acp{dt}.
    \item There is a need to develop efficient yet, scalable scheduling mechanisms to realize massive access control, in an ad-hoc, distributed fashion.
\end{itemize}  & \begin{itemize}
    \item Challenges due to massive number of nodes requiring simultaneous communication with their respective twins.
    \item Increased inter- and intra-twin interference, and hence, reduced \ac{qos}.
\end{itemize} \\ \midrule
Reliability & \begin{itemize}
    \item Ultra-high reliable communication is essential to ensure accurate models updates exchange.
    \item Has a profound impact on the global models shared among distributed \acp{dt}, and accordingly, has serious consequences on the Metaverse operations.
\end{itemize} & \begin{itemize}
    \item Extreme reliability requirement is a consequence of the need to perform immersive, high-resolution multi-media streaming between the cyber and physical twins.
\end{itemize}   \\ \midrule
Latency & \begin{itemize}
    \item Ultra-low latency is required to ensure on-demand multimedia streaming and full-immersive experience.
\end{itemize} & \begin{itemize}
    \item Latency requirements in intra-twin communication is alleviated
\end{itemize} \\ \midrule
Throughput &  \begin{itemize}
    \item Inter-twin communication does not demand extremely high rates, compared to intra-twin communication, i.e., rate levels need to satisfy a predefined thresholds, in order to ensure the efficient transmission of the models updates.
\end{itemize} & \begin{itemize}
    \item The immersive nature of the data communicated between the digital and physical twins necessitates the design of new high-rate communication paradigms, that satisfy the energy and latency requirements of intra-twin links.
\end{itemize} \\    \midrule 
\end{tabular}
\end{adjustbox}
\begin{adjustbox}{max width=\textwidth} 
\begin{tabular}{c|c|c|c|c|c|c}
    
     {\bf Peak data rate } &  { \bf \acs{e2e} latency} & { \bf Network density}  & { \bf Peak spectral efficiency}   &   { \bf Receiver sensitivity } & { \bf Positioning precision } & { \bf Reliability } \\ \midrule
    1 Tbps  & $<$ 1 ms &  $>$ 1M device/$km^2$  &  
  60 b/s/Hz &  $>$ -130 dbm &
           Centimeter level & 99.9999$\%$ \\ 
        \bottomrule
    \end{tabular}
\end{adjustbox}
\vspace{-3mm}
\end{table*}

{\bf ISO/IEC 23090-10}. The storage and carriage of \ac{v3c} data are specified by the \acs{mpeg} system ISO/IEC 23090-10 standard. Like \ac{omaf}, ISO/IEC 23090-10 leverages existing system standards designed for 2D video. More specifically, the ISO/IEC 23090-10 standard defines how \ac{v3c} data is stored on \ac{isobmff} containers and specifies extensions to the \ac{dash} for delivery over the network. The standard defines three ways for storing \ac{v3c} data on \ac{isobmff} (ISO/IEC 14496-12) containers, including single-track storage, multi-track storage, and non-timed storage. This latter enables the storage of static \ac{v3c} objects, which can also be used as thumbnails of the volumetric video track. The single-track storage enables carriage of \ac{v3c} data on an \ac{isobmff} container with limited functionalities. In contrast, multi-track storage encapsulates \ac{v3c} components on different tracks enabling advanced features such as preventing bitstream demultiplexing prior decoding. In addition to \ac{isobmff} boxes, the  ISO/IEC 23090-10 standard introduces new boxes specifically for \ac{v3c} data, such as \ac{v3c} decoder configuration box and unit header box. This latter stores four bytes \ac{v3c} unit header used to identify \ac{v3c} components properly and maps them to the active \ac{vps} signaled in the \ac{v3c} decoder configuration box. 

For distribution, the ISO/IEC 23090-10 specification supports the live streaming of \ac{v3c} content based on the \acs{mpeg}-\ac{dash}. Furthermore, the standard specifies how \ac{v3c} segments are signaled in the \ac{mpd} in both single and multi-track encapsulations. In the former configuration, the \ac{v3c} content is described in the \ac{mpd} by a single adaptation set with one or multiple representations encoded by the same codec. On the other hand, the multi-track encapsulation offers more flexibility since each \ac{v3c} component is represented by an adaptation set. Therefore, this representation allows the client to perform adaptive streaming by requesting only a set of components. In addition, the adaptation sets can be encoded with different codecs or bitrates, enabling adaptive bitrate streaming.    



\textcolor{black}{With the different types of multimedia data and transport protocols, that are essential for the development of an immersive Metaverse, it is essential to ensure that future wireless networks can provide resilient and ultra-low latency multimedia transmission between the cyber and physical realms. Therefore, in the following section, we articulate multiple wireless technologies that we envision will be key pillars in the Metaverse paradigm.}

\begin{figure*}[h!]
\centering
\includegraphics[width=0.85\textwidth]{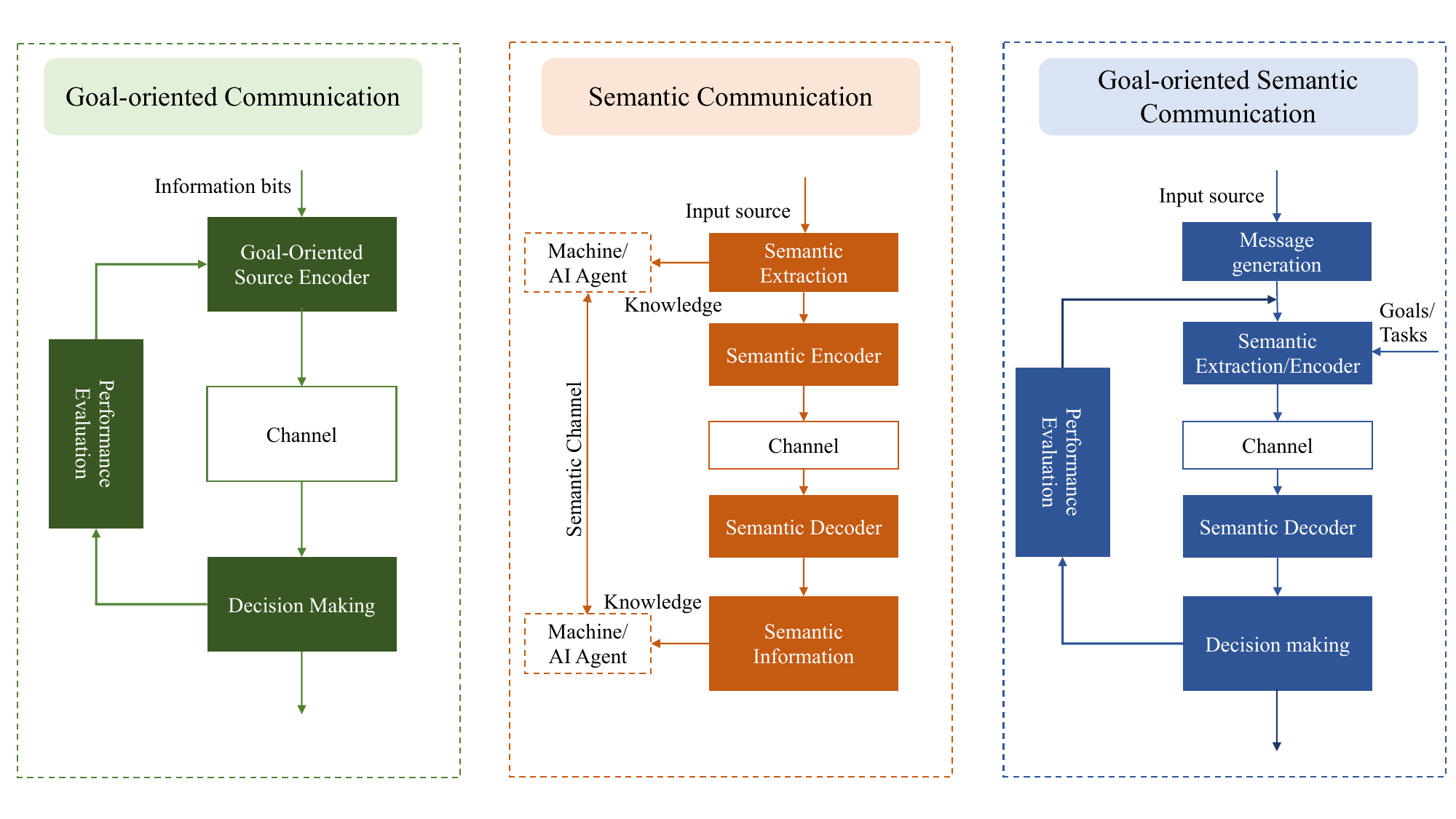}
  \caption{Architecture of goal-oriented, semantic, and goal-oriented semantic communication.}
  \label{fig:SC-GO} 
  \vspace{-5mm}
\end{figure*}
\section{6G: Emerging Technologies for Immersive Media Communication}
\label{sec:meth}

The ultimate goal of realizing an immersive, interoperable, and holistic Metaverse can be accomplished through enabling the massive twinning paradigm, in which a large number of distributed \acp{dt} are linked in order to create an inclusive, comprehensive digital realm that comprises all objects-related details and functionalities of its physical counterpart. {It is worthy to highlight that the use of a massive number of digital twins (which form the Metaverse) could enhance the fidelity of this virtual representation, enabling more realistic and immersive monitoring and interactions within the Metaverse. To achieve this, multimedia data (incorporating videos, images, sound signals, etc.) will need to be continuously communicated between the cyber and physical twins, to ensure a synchronized, up-to-date, and accurate Metaverse. Within this context, the role of 6G becomes essential, where a reliable and fast communication will be needed to ensure a smooth connection between these twins and therefore, a homogeneous Metaverse, imposing particular requirements on future wireless generations. Current communication schemes are incapable of delivering the needed level of immersion when communicating real-time multimedia data, and hence, there is a need to revisit existing scheme and develop novel techniques for improved communication through 6G.} 

 In more details, each \ac{dt} is envisioned to model, monitor, and control a single (or a multiple) network segment/s or function/s, and ideally, a massive twinning paradigm will pave the way to a global \ac{dt} that comprises all network nodes and their interactions with each other and with the environment \cite{uusitalo20216g}. Accordingly, the inter-twin communication concept emerges as a mean to bridge multiple twins, which refers to the communications happening at servers (cloud- or edge-servers) between two or more twins. Inter-twin communication aims to allow sharing of models, knowledge, and feedback among the twins and, subsequently, contribute to further reducing the overhead from available computing resources, as will be discussed later in this section. Another essential communication paradigm, intra-twin communication, is required to ensure a real-time operational \ac{dt} through reliable links between the physical and cyber twins. Although both communication paradigms share some challenging requirements, each has unique demands attributed to the distinct goals associated with each paradigm. While intra-twin communication necessitates powerful computing capabilities and high capacity, the performance of inter-twin communication is characterized by the links' reliability, rate, and latency. In Table \ref{Tab:inter-intra}, we highlight the key requirements of each communication paradigm, and network requirement in order to achieve reliable media communication \cite{bariah2023ai}. It is worth highlighting that distributed \acp{dt} can be placed at cloud servers, edge serves, or a combination of both. In this section, we will shed lights on the main communication paradigms that are identified as the key to enabling immersive multimedia communication in the \ac{dt} and the Metaverse. Given that intra-twin communication is aimed at models' parameters exchange and knowledge transfer, we will focus on enabling paradigms of inter-twin communication, where the latter is aimed at immersive media transfer between the physical and cyber twins.

\subsection{Beyond Shannon Communication}

Future wireless generations are envisioned to enable more intelligent services, particularly concerning machines/\ac{ai} agents communications, while coping with the available network resources, including energy, spectrum, and computing resources. We strongly believe such a vision cannot be realized solely by relying on higher frequency bands or conventional energy-efficient mechanisms. Within this context, upcoming wireless generations are anticipated to allow machines to communicate meaningfully while satisfying particular rate and energy constraints, by extracting the useful semantics to be communicated with the receiver instead of transmitting the whole message (which lacks in most scenarios the semantic aspect that allows the receiver to understand the purpose/meaning of the message). It is further envisaged that machines will exploit extracted semantics in order to perform particular predefined goals pertinent to parameters' estimation, optimization, classification, etc., in order to allow self-optimizing and self-configuring networks. Motivated by this, in this subsection, we shed lights on the essential role of goal-oriented, semantic, and goal-oriented semantic communication paradigms (See Fig.~\ref{fig:SC-GO}) in enabling immersive multimedia communications. 

\subsubsection{Goal-Oriented Communication}
Multimedia data communication can be effectively achieved by establishing particular goals to be fulfilled by the interacting nodes between the physical environment and the \ac{dt}, in which these goals impose specific requirements pertinent to latency, resolution, field-of-view, etc. This scheme is regarded as goal-oriented communication. By defining a set of goals to be achieved when communicating particular media, the system specifications and constraints are then linked to this set, and interacting entities will aim to deliver the \acp{kpi} with relevance to the identified goals. In goal-oriented communication, communication overhead is reduced through dedicating network resources to transmitting only essential information that directly contributes to the fulfillment of the goals' objectives. Meanwhile, information that doesn't have implications on the defined set of goals is considered less significant and, therefore, will be neglected in this process \cite{strinati20216g}. \\
The key concept of goal-oriented communication in multimedia data is that transmitting nodes at the physical twin put more focus on reliably transferring data related to particular features within the user's field-of-view, thereby readily affecting the quality of the compression and contributing to the spectral efficiency enhancement of multimedia communication between the physical and cyber twins. In \cite{zhang2022goal}, it was demonstrated that goal-oriented communication could be effectively applied to deploy data compression for high-dimensional data, including goal-oriented signal processing, data quantization, and clustering. However, one of the common limitations facing goal-oriented communication within the \ac{dt} paradigm is the high possibility of experiencing heterogeneous tasks with diverse requirements, rendering network re-configuration a time-consuming and challenging process. Within this regard, \ac{ai}-enabled solutions should be designed for more generalized and adaptive goals, with the aim to fulfill the requirements of multiple tasks within the network, given unified network setup and configuration.  

\subsubsection{Semantic Communications}
As demonstrated earlier, wireless communications in the \ac{dt} paradigm are anticipated to support distributed multimedia exchange, particularly continuous multi-modal media such as graphical animations, high-quality audio and video, haptics, and interactive images. Such real-time multimedia communication demands a large bandwidth in order to deliver the promised high \ac{qoi}. Semantic communication alleviates the need for highly reliable and low-latency wireless channels, and therefore, real-time multimedia streaming with optimum synchronization between the cyber and physical twins can be achieved by relying on the available computing resources \cite{chaccour2022less}. 

Although immersive multimedia data are multi-modal in nature, there should be a sort of correlation among different modalities. Accordingly, the role of semantic communication is not limited to extracting the unique meanings from each data stream but also understanding cross-modality features in different multimedia signals and then communicating the cross-modal semantics with the \ac{dt} servers \cite{li2022cross}. Such an approach can further reduce the bandwidth needed to stream multimedia signals with the cloud while ensuring an improved \ac{qoe} for cross-modal applications at the \ac{dt} network. Nevertheless, it is essential to quantify to what extent cross-modal semantic communication can deliver the needed immersive experience while satisfying particular rate requirements. Further, it was demonstrated that cross-modal semantics could help overcome the polysemy and ambiguity problems experienced in single-modal scenarios, primarily due to a lack of understanding of the context from which the semantics are extracted. Accordingly, efficient semantic communication systems constitute a promising approach toward extracting cross-modal features that interpret the needed information within the overall context, thereby introducing enhanced throughput and reliability performance. 

\subsubsection{The Interplay of Goal- \& Semantic-Aware Communication}

Driven by their inherent nature of communicating an abstracted version of wireless messages, the intertwined advantages of goal-oriented and semantic communications can be ultimately combined with the aim to achieve reliable, yet spectrally efficient multimedia communication. From one perspective, network goals can be set in order to maximize the effectiveness of the transmitted semantics, and thereby, enhancing the reliability performance of semantic communication systems. On the other hand, semantic communication plays an important role in further enhancing the throughput performance of goal-oriented communication, in which semantic encoder/decoder are employed in order to extract the useful information that are necessary for the achievement of the intended network goals, and hence, enhance the overall spectral efficiency of the network.

\subsection{Communication Protocols}
Supporting multimedia-on-demand traffic necessitates revisiting current communication protocols in order to design new protocols that can strike a balance between \ac{qoi}, energy, latency, and rate. It is recalled that classical \ac{mac} protocols are generic and uninterpretable, and their control signaling messages are unadaptable and hence, cannot be optimized for a particular task in order to deliver a task-specific \ac{qos} \cite{seo2022towards}. With the diverse requirements of different tasks within the \ac{dt} paradigm, efficient emerging communication protocols should be designed to handle the large volume of multimedia data to be shared over the wireless medium in a latency- and reliability-sensitive manner. Emphasizing on this, conventional approaches rely on granting early slots for high-priority data to guarantee a particular latency threshold, while reliability is ensured through redundancy and re-transmission schemes. However, within multimedia immersive communication for \ac{dt}, such approaches are not well-suited for provisioning the needed \ac{qoi} in terms of delay and reliability. In addition, the definition of high priority and reliability is different when tackling multi-modal data, where each stream, and each packet within the stream, has different \ac{qos} requirements. For instance, a packet with I-frames from video data has higher latency and reliability requirements than packets with P- or B-frames.  \\
Therefore, a cross-design between multiple layers defines the new concept of multimedia communication, where data encoding, compression, and communication are performed through learned goal-oriented protocols, which are optimized to achieve identified \ac{qos} requirements. Within this context, \ac{ai} based protocols have emerged to realize optimum signaling and medium access policies. \Ac{marl} has demonstrated a promising reliability performance when \acp{ue} employ \ac{marl} to learn how to communicate without {a prior} knowledge of the \ac{mac} protocol \cite{hoydis2021toward}. In specific, such an approach enables interpretable protocols, in which receivers can interpret various information related to control messages, timing advance, power headroom report, buffer status report, etc. It is envisioned that \ac{ai}-based communication protocols will introduce a tangible performance enhancement for multimedia communication through realizing overhead reduction and medium access coordination optimization, whether \ac{ai} agents are utilized for allowing learning-to-communicate through standardized protocols or as a paradigm shift towards more innovative protocols that are designed by \ac{ai}. 

\subsection{Holographic-Type Communication}
The future vision of \ac{htc}, that is capable of providing the required level of immersion at the Metaverse, is to realize an efficient integration between 3D capturing, hologram encoding, compression and generation, and 3D hologram transportation and display in sub-millisecond latency and ultra-high frame-rate. In particular, such technology is anticipated to allow network users to enjoy a full-sense interactive experience with the \ac{dt} of interest and the Metaverse. Ensuring high-definition holograms necessitates that the acquired data at the capturing stage be inclusive to all senses, and these senses should be transported in a synchronized manner. This is particularly for vision-based data. Accordingly, a tracker is employed to continuously ensure synchronized streams among movements, gestures, and other visual data \cite{petkova2022challenges}. 

\Ac{htc} relies mainly on depth sensors (i.e., LiDAR 3D cameras) and \ac{ar} services to guarantee reliable communication of 3D holograms with parallax and haptics featuring \ac{6dof} (incorporating translation and rotation movements), with the aim to deliver the needed sensory experience \cite{Ericsson-HTC}. However, in order to enable high-fidelity holograms in the Metaverse, ultra-high bandwidth (Gbps) and close-to-zero delay between the physical object and its hologram constitute key limiting factors. The streaming overhead of holograms can be reduced by developing sophisticated prioritization mechanisms in which multiple streams are prioritized according to their impact on the user's perception. Specifically, several metrics, including frame-rate, angular resolution, depth, dynamic range, etc., can be optimized in an adaptive fashion, taking into account the field-of-view of the user. In this way, resources are allocated to streams and frames that contribute to enhancing the hologram's \ac{qoi} \cite{clemm2020toward}.  

\begin{figure}[t]
\centering
\includegraphics[width=0.48\textwidth]{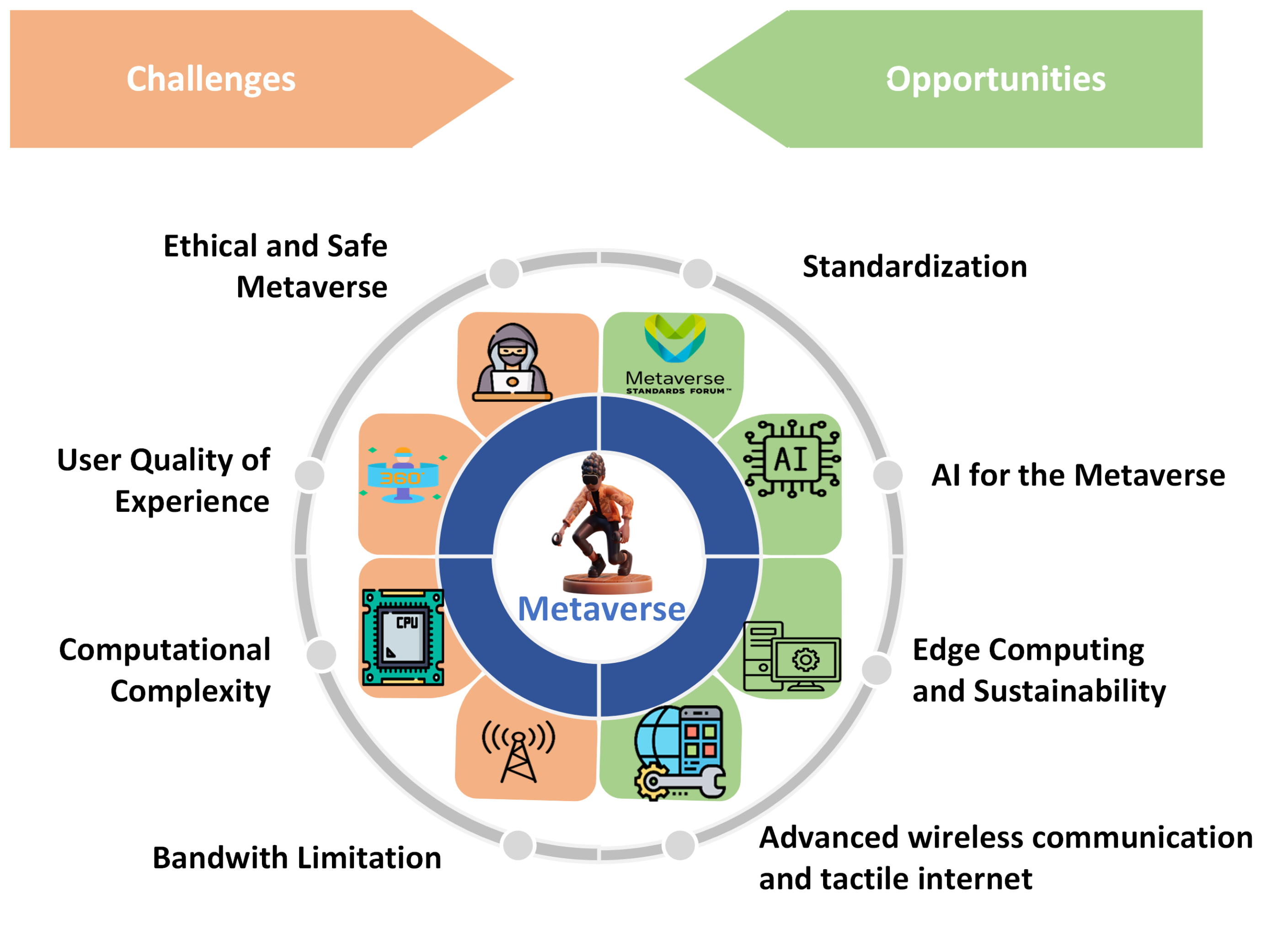}
  \caption{{\color{black} Main challenges and opportunities for the development of the Metaverse.}}
  \label{fig:chal-meta} 
  \vspace{-5mm}
\end{figure}

\section{Open Challenges \& Future Research Directions}
\label{sec:opencha}
{\color{black}The complete integration of the Metaverse with the real world is imperative for its comprehensive realization. This endeavor, however, is fraught with technical challenges, as outlined in Fig~\ref{fig:chal-meta}. These challenges span diverse domains, including user ethics and security, \ac{qoe} for users, interoperability, computational complexity, and constraints related to bandwidth in wireless multimedia communication. To address these challenges effectively, this section explores various opportunities, leveraging strategies such as standardization, the integration of \ac{ai} in the Metaverse, cloud and mobile edge computing, as well as advancements in network communication and tactile internet.} \vspace{-3mm}
\subsection{{\color{black} Standardization}}
Several standardization bodies have explored 3D visual content at different levels of the transmission chain, from coding and streaming at the ISO/IEC \acs{mpeg} to transmission over 5G networks and broadcast at \acs{3gpp} and \ac{dvb}, respectively. These standards provide tools for efficiently streaming 3D visual content while ensuring the interoperability of devices and fast deployment of immersive video services. Still, the ISO/IEC \acs{mpeg} 3D graphics group launched in 2021 a call for proposals for a new 3D dynamic mesh coding standard called \ac{v-dmc} (ISO/IEC 23090-29)~\cite{9922888}, as a new member of the \ac{v3c} standards family. {The \ac{v-dmc} standard is currently in development, with the goal of achieving efficient compression for both lossless and lossy scenarios involving 3D dynamic meshes. The applications targeted by this standard include real-time 3D video streaming, rendering, low-latency cloud gaming, and more. Further, a coding solution based on \ac{nerf} cannot be deployed on existing systems due to the lack of interoperability in the model weights stream and the architecture of the neural radiance field model. Standardization activities are necessary to define a standard representation for weights in the bitstream and the architecture of the model used by end-users to reconstruct the static or dynamic 3D scenes on different devices. The file format of the bitstream generated by the encoder should also be standardized to enable rendering on existing video players with streaming features. This includes support for random access points and the seamless playback of dynamic objects. These standard neural codecs and streaming protocols can then be integrated into the 3GPP specifications to be supported by the next-generation of wireless communication standards. }

Furthermore, the \ac{3gpp} has worked on several projects related to \ac{vr} and \ac{xr}. For instance, the \ac{3gpp} technical report 26.918 attempts to identify interoperability issues that may require standardization activities in the \ac{3gpp} context with a focus on \ac{odv} and associated audio. Further, the \ac{3gpp} technical specification 26.118 defines media and representation profiles for the interoperability of \ac{vr} streaming services. Moreover, the \ac{3gpp} technical report 26.928 collects information on \ac{xr} in the context of 5G radio and network services to identify potential needs for \ac{3gpp} standardization. In addition, the \ac{vrif} guidelines\footnote{VR industry forum guidelines v2.3: \url{https://www.vr-if.org/wp-content/uploads/vrif2020.180.00-Guidelines-2.3_clean..pdf}} provide practices and recommendations for high-quality and interoperable 360$^\circ$ video services. More recently, the \ac{vrif} has released a guidelines document\footnote{Volumetric video industry forum guidelines v1.0: \url{https://www.vr-if.org/wp-content/uploads/Volumetric-Video-Guidelines-1.0.pdf}} covering all aspects of volumetric video delivery ecosystem targeting high-quality \ac{vr}, \ac{ar} and \ac{xr} services. Finally, the ITU-T launched on December 2022 the \ac{fg-mg} under the telecommunication standardization advisory group. This group will investigate the technical requirements of the Metaverse to identify the fundamental enabling technologies from multimedia and network optimization to digital currencies, \ac{iot}, \acp{dt}, and environmental sustainability\footnote{\Acl{fg-mg}: \url{https://www.itu.int/en/ITU-T/focusgroups/mv/Pages/default.aspx}}. For instance, the \ac{usd} NVIDIA open software provides a rich, common language for defining, packaging, assembling, and editing 3D data. Therefore, \ac{usd} is a good candidate to become a standard facilitating the use of multiple digital content creation applications for their integration in Metaverse platforms.   

\color{black}
\subsection{{\color{black} AI for the Metaverse}}
\noindent {\bf {\color{black} Conventional AI.}} The breakthrough advancements in \ac{ai} pave the way for exploiting various \ac{ai} models for improved \ac{qoe} of Metaverse users. In particular, \ac{nerf} \cite{10.1007/978-3-030-58452-8_24} represents an innovative \ac{ai} concept for the Metaverse. \Ac{nerf} has emerged as an \ac{ai}-based solution that implicitly encodes the radiance field of the 3D scene in a \ac{mlp} network. This \ac{mlp} takes as input continuous 5D coordinates and predicts the volume density and view-dependent emitted radiance at the input spatial location. The \ac{nerf} model trained on a sparse set of input views achieves state-of-the-art performance for synthesizing a novel view through a continuous representation of the scene. The success of \ac{nerf} has revolutionized the 3D representation and rendering based on a lightweight \ac{mlp} network. Several papers have then been published addressing \ac{nerf} shortcomings: 1) reducing the memory access and computational complexity of the training, 2) real-time rendering on mobile devices, and 3) extension to 3D dynamic scenes. Therefore, the recent developments of \ac{nerf} will shape the Metaverse with compact representation and real-time rendering of static and dynamic 3D scenes in high fidelity. \\\\ 
{\bf {\color{black} Generative AI empowered Metaverse.}}
The recent advancements in generative \ac{ai}, and \acp{llm}, show a promising potential to reshape the different types of experiences within the Metaverse. The main benefits of leveraging \acp{llm} in the Metaverse is their capabilities to perform on-demand content generation, in a versatile and adaptive manner, fulfilling the needs of several scenarios within the Metaverse. This is particularly pronounced with the emergence of visual generative models, e.g., DALLE-2 (text-to-image), Phenaki (text-to-video), and DreamFusion (text-to-3D image), which constitute a clear path towards generating high-fidelity visual content, while ensuring efficient bandwidth usage, lower latency, and enhanced functionalities. From a different perspective, ChatGPT3/4  is a powerful conversational system released recently by OpenIA\footnote{\url{https://openai.com/blog/chatgpt/}}, based on the \ac{gpt3} model. The outstanding conversational capabilities of ChatGPT3/4 make it a promising candidate to provide the Metaverse with many compelling features in high fidelity. Integrating such a model with intelligent agents will enable more natural and immersive interactions with the virtual world through understanding and responding to human language. Furthermore, the ChatGPT3/4 model will allow personalized experience through fine-tuning the model to a specific application or language. Therefore, generative \ac{ai} models will offer the Metaverse a more natural way of communication, collaboration, and creativity, and thereby, enhancing the user's \ac{qoe}. 
\subsection{{\color{black} Edge computing and sustainability}} 
\noindent {\bf Cloud and mobile-edge computing.} Realizing massive twinning at cloud or edge servers impose high computing overhead on cloud and edge resources, particularly, to process multimedia data. In particular, computing entities in the Metaverse paradigm is expected to process a vast amount of high-dimensional data in order to allow the user to experience the full immersion in the virtual world. In addition to the data volume, novel challenges pertinent to data 3D rendering, which rely on converting raw multimedia data into displayable objects, has emerged with the Metaverse paradigm, demanding further computing resources from the \ac{dt} servers. Accordingly, it is necessary to develop efficient multimedia computing services, not only for Metaverse providers but also for end users who wish to access the Metaverse through their resource-constrained devices. This includes effective task-offloading and 3D rendering approaches. Several \acp{npu} available on embedded devices can run \ac{ai} models in real-time and with low energy consumption. For example, the \ac{npu} integrated on the latest Snapdragon Qualcomm mobile devices enables up to 32 Tera operations per second in 8-int, 16-int or 32-float precision. In addition, edge computing devices such as the NVIDIA Nano board enables up to 472 G\ac{flops} in 16-float precision with 128 cores. \\\\
{\bf Sustainable Metaverse.} The Metaverse will drive sustainability by reducing several applications' energy consumption and \ac{co2} emissions. Specifically, the Metaverse will shift consumer and industrial applications by moving from harmful physical mobility towards more sustainable interactions in digital spaces. Nevertheless, the Metaverse also raises serious concerns about its software, hardware, and infrastructures that need to be eco-friendly and designed to minimize their \ac{co2} footprint. For instance, \ac{gpt3} model has 175 billion parameters, and its training energy footprint is estimated to 936MWh, which corresponds to the energy consumption of 30,632 American householders or 97,396 European householders. The \ac{ai} models are becoming larger and larger to handle more complex tasks and reach the level of intelligence and adaptation of the human brain. Nevertheless, our brains are energy efficient, consuming less than 40 Watts for 100 peta\ac{flops} computing power, which is faster than any existing supercomputer. Therefore, future research can get inspired by the human brain to perform efficient and sustainable computing of large \ac{ai} models within the Metaverse platform. {\color{black} Alternative short-term strategies involve the development of more efficient generative models with a reduced number of parameters. By prioritizing model optimization and utilizing high-quality data, it becomes feasible to achieve performance levels comparable to larger models while significantly reducing energy consumption. Model optimization encompasses techniques such as distillation, pruning, and quantization, aiming to create lightweight models that enhance efficiency and minimize environmental impact.}

\subsection{{\color{black} Advanced wireless communication and tactile internet}}
\noindent {\bf Theoretical limits.} The prevailing vision of future wireless communication is directly attached to breaking the Shannon-limit barrier, in which semantics and goal-oriented communications are anticipated to be the basis of multimedia data communication, and hence, the key driver behind the Metaverse, as discussed earlier in this article. Accordingly, extremely higher data rates, with tight energy constraints, are envisioned to be realized in future 6G networks. However, it is yet unclear to what extent we can go beyond the Shannon limit. Specifically, there are no theoretical underpinnings that can quantify the performance of such technologies and validate the hypothetical claims on the achievable rate by semantic- and goal-oriented communications. This is further pronounced in the \ac{dt} and the Metaverse paradigms, which lack a solid mathematical representation, and therefore, the fundamental limits of multimedia communications in massive twinning through semantics or goals constitute an open research topic. \\\\
{\bf Tactile internet.} The Metaverse is envisioned to provide a full-immersive experience to Metaverse users through the real-time exchange of the five senses, including the haptics. Tactile internet, as an enabler for haptic communication, enables real-time gathering, perceiving, and controlling virtual objects according to haptic feedback. Haptic feedback in the Metaverse is expected to offer kinesthetic and tactile perception to efficiently complement other senses towards extremely-high \ac{qoi}. However, several challenges should be tackled prior to the real adoption of tactile internet in the Metaverse, including reliable haptic interfaces, joint communication and control for teleportation, and user privacy.  
\subsection{{\color{black} Ethical \& safe Metaverse}}
The future vision of the Metaverse relies on exploiting a massive number of sensors that are attached to everything in the physical environment in order to enable two-way control between the \ac{ct} and \ac{pt}. Such collected data may contain highly sensitive and private data, including biometric data, and therefore, it is anticipated that the Metaverse will be an open world subject to several attacks, compromising users’ privacy and safety. This includes the risk experienced through nodes/users interactions inside the Metaverse, which is caused by unethical behaviors conducted by Metaverse users. This issue is exacerbated by the employment of immersive media, which renders a wide range of details within the communicated data. Accordingly, activities pertinent to ensuring effective fulfillment of the privacy policies, regulations, ethical conducts, and users' accessibility of the Metaverse are essential to be initiated at the early stage of the development of the Metaverse paradigm.  
\subsection{{\color{black} User quality of experience}} 
\noindent {\bf Rendering 3D visual content.} 
Rendering volumetric video content on mobile devices with constrained computing and energy resources is still challenging, especially when multiple volumetric objects are present in the scene, in addition to the interaction of the user with the scene. Remote or interactive rendering reduces the processing load on the user and enables efficient bandwidth usage by sending only a 2D view of the volumetric object according to the user's position. However, this solution increases both the motion-to-photon latency and the glass-to-glass latency requiring additional transcoding in the cloud. To minimize the latency, the rendering can be performed in a geographic area close to the client by leveraging \ac{mec} along with the low latency and high throughput offered by 5G/6G networks and the WebRTC communication protocol. In addition, accurate user pose prediction algorithms can be used to reduce the latency further and increase the perceived quality, preventing users' discomfort for a better viewing \ac{qoe}. {In addition, real-time rendering of \ac{nerf}-based solutions on embedded devices with low computing and memory resources remains an open challenge. Several solutions can be investigated to first create, from pre-trained \ac{nerf}, a mesh representation that can be efficiently processed by standard graphics pipelines, available on wide range of resource-constrained devices, including mobiles and AR/VR headsets. Finally, the advancement of hardware decoders for \ac{miv}, V-\ac{pcc}, and G-\ac{pcc} standards is imperative to facilitate their deployment on mobile and wearable devices. This enhancement would enable real-time decoding capabilities and contribute to low energy consumption.} \\\\ 
{\bf Assess the users \ac{qoe}.} The success of the Metaverse is mainly related to users' satisfaction with the experience provided by Metaverse platforms. Several criteria can play an essential role in the final \ac{qoe}, including audio/video quality, \ac{e2e} or photon-to-motion latency, computing resources, network throughput/latency/jitter, and finally, the viewing comfort offered by display devices. All these parameters need to be jointly optimized to reach the high comfort, and the \ac{qoi} promised today by the Metaverse. Metaverse platforms also challenge the quality estimation research community to build new accurate models that estimate the user \ac{qoe} in Metaverse platforms. In contrast to traditional image/video quality estimation models, additional features of the Metaverse platforms, network, and user feedback must be considered to build new datasets and accurate \ac{qoi}/\ac{qoe} models.\\\\
{\color{black} \bf Towards advanced 3D media modality.}
Digital holography (DH) can represent the 3D information of the scene through wavefronts of light. The digital holography acquisition system captures the wavefronts' phase and amplitude optically with a digital sensor array or numerically using hologram synthesis algorithms. The resulting patterns are processed and stored as \acp{dh}. Then, the \ac{slm} technology is used to represent the \acp{dh} in optical setups for displaying the 3D scene in the air without requiring a physical display. Nevertheless, the resolution supported by the available \ac{slm} is insufficient to display most of \acp{dh} in high-quality. This latter issue is addressed by viewing the \ac{dh} on legacy 2D displays. To reach this end, numerical models are used for reversing light propagation and then post-processing the complex-valued floating-point wavefield to obtain regular images. Multiple propagation models and associated pre/post-processing modules have been developed in the literature. In particular, the ISO/JPEG Pleno (ISO/IEC 21794-1...6) efforts aim to standardize coding tools for 3D static visual modalities, including light field (ISO/IEC 21794-1), point clouds (ISO/IEC 21794-6), and \aclp{dh} (ISO/IEC 21794-5). JPEG Pleno defined in the \acp{ctc}  a new \ac{dh} images dataset and the \ac{nrsh}. This latter is a Matlab software that enables extracting specific views from the Pleno \ac{dh} dataset, providing accurate reconstructions. Nevertheless, in addition to the high computational complexity of holography reconstruction algorithms, their main limitation is the low-quality reconstructed views due to severe aliasing and speckle noise distortions, preventing their widespread deployment. 

\section{Conclusion}
\label{sec:con}
In this article, we have approached the development of the Metaverse from the angle of immersive multimedia and the enabling wireless technologies that support the smooth and reliable communication of immersive media between the physical and digital worlds. Specifically, we thoroughly discussed the different media modalities, highlighting their pros and cons, standardization activities, and coding techniques. We further explored the potential of 6G networks in realizing efficient transmission of the surveyed multimedia data, where we have outlined the key wireless technologies that are envisioned to be the Metaverse underpinnings. We finally sketch the roadmap towards developing an efficient Metaverse by outlining the main limitations of wireless multimedia communication and opening new horizons of future research directions.

\bibliographystyle{IEEEtran}
\bibliography{ref}

\section*{Biographies}

\textbf{Wassim Hamidouche} (Wassim.Hamidouche@tii.ae) is a Principal Researcher at the Technology Innovation Institute in Abu Dhabi, UAE. 

\textbf{Lina Bariah} (lina.bariah@ieee.org) is a Lead AI Scientist at Open Innovation AI in Abu Dhabi, and Adjunct Professor at Khalifa University, Abu Dhabi.  

\textbf{M{\'e}rouane Debbah} (Merouane.Debbah@ku.ac.ae) is a Professor and Director of KU 6G Research Center, Khalifa University, Abu Dhabi.

\end{document}